# Ho:YLF amplifier with Ti:Sapphire frontend for pumping mid-infrared optical parametric amplifier

Krishna Murari,[1,*] Fangjie Zhou[1], Yanchun Yin,[1] Yi Wu[1], Bruce Weaver[2], Timur Avni[2], Esben Larsen[2], Zenghu Chang[1]

[1]*Institute for the Frontier of Attosecond Science and Technology, The College of Optics and Photonics (CREOL) and Department of Physics, University of Central Florida, Orlando, FL, 32816*
[2] *Quantum Optics and Laser Science Group, Blackett Laboratory, Imperial College London, London, SW7 2BW, UK*
*\*Corresponding author: krishnacelos@gmail.com*

**We present a Ho:YLF Chirped-Pulse Amplification (CPA) laser for pumping a longwave infrared Optical Parametric Chirped Pulse Amplifier (OPCPA) at a 1 kHz repetition rate. By utilizing a Ti:Sapphire laser as a frontend, 5-μJ seed pulses at 2051 nm laser pulse are generated in a Dual-Chirp Optical Parametric Amplifier (DC-OPA), which are amplified to 28 mJ pulses with a pulse duration of 6.8 ps. The scheme offers a potential driver for two-color (800 nm and 8 μm) high harmonic generation with an increased keV X-ray photon flux.**

*OCIS codes:*

Ultrashort intense mid-wave and long-wave-infrared (MWIR, LWIR) laser pulses are of great importance to strong-field physics and attosecond science experiments. Recent development of carrier-envelope phase (CEP)-stabilized few-cycle lasers at a wavelength of 1.6 – 2.1 μm have paved the way for the next generation of attosecond light sources with photon energies reaching the water window (284 eV – 530 eV) [1,2]. Scaling of photon energies by high-order harmonic generation (HHG) to a few keVs requires the development of high-energy few-cycle pulses extending into the long-wave infrared wavelength because the cut-off photon energy of the HHG process scales quadratically with the driving laser wavelength. Furthermore, high energy LWIR lasers are an ideal tool to study HHG in solids [3], acceleration of electrons in dielectric structures & plasma [4], breakdown of the dipole approximation [5] and rotational/vibrational spectroscopy [6]. X-rays at 1.6 keV have been produced using lasers at 3.9 μm [7]. However, the conversion efficiency decreases with the increasing wavelength of the driving laser [8], thereby reducing the total flux of the attosecond pulses. To be able to perform time-resolved pump-probe experiments such as transient absorption spectroscopy and attosecond streaking, innovative schemes are needed to enhance the attosecond photon flux when driven with LWIR lasers. In this context, multi-color laser fields have been suggested to increase the harmonic yield and extend the cut-off energy [9]. For instance, when the HHG process is driven by two-color fields (a strong IR field and a weak attosecond extreme ultraviolet) the harmonic yield can be enhanced [10]. In the case of keV X-rays driven by a LWIR laser, it has been predicted theoretically that the intensity of an X-ray pulse can be significantly enhanced by adding a few times weaker few-cycle CEP-stable near infrared (NIR) pulse centered at 800 nm [11].

High-energy femtosecond MIR/LWIR pulses can be generated using an optical parametric chirp pulse amplifier or an optical parametric amplifier (OPA) involving frequency conversion in a nonlinear medium. Most lasers developed so far in this regime are limited to wavelengths up to 3 - 4 μm due to the oxide-based nonlinear crystals used, which have limited transparency beyond this wavelength. Extending the wavelength further into the LWIR requires the frequency conversion in non-oxide crystal like zinc germanium phosphide (ZGP). However, such crystal is not transparent below 1.9 μm, therefore pumping the OPA at longer wavelengths than provided by ytterbium (~1.03 μm) and neodymium (~1.06 μm) doped YAG lasers are required. Hence, the development of high-energy longer wavelength (>1.9 μm) pump sources becomes crucial in the generation of high peak power LWIR pulses. In the recent years, several works have been published in this regard generating a wide range of pulse energies in the 2-μm wavelength range based on rare earth Ho[+3] (holmium) ion doped gain medium [12–15]. In this regard most of the 2-μm high energy lasers are based on Ho:YLF or Ho:YAG developed so far. In addition they are seeded by nJ-level pulses from fiber lasers and hence required high-gain amplification typically possible through 30-40 round-trips in a regenerative amplifier (RA).

Grafenstein et al. utilized a Ho:YLF based RA to reach 10 mJ, which was then amplified in a two-stage booster linear amplifiers to reach an energy up to 55 mJ. In this case, an amplified gain of $10^8$ was achieved owing to the typical damage threshold of the Ho:YLF crystal of 16 GW/cm$^2$. The high gain in the system leads to a long output pulse duration owing to gain narrowing. Nevertheless, by utilizing this system, an OPCPA emitting 1 mJ, 100 fs pulses at 5.1 µm has been demonstrated [16]. However, in order to use this system to generate sufficient HHG yield for pump probe experiments, as mentioned before, it would be beneficial to add a second NIR pulse at, for example, 800 nm for two-color HHG scheme. Such experiments require mJ-level single-cycle pulse at both wavelengths with CEP stability. The CEP stability becomes a challenge if the Ho:YLF pump laser is seeded by a different laser as that of the seed generated for the LWIR OPCPA. Hence, in order to perform two-color HHG it is advantageous to use a Ti:Sapphire as a front-end which can serve as a seed for three-fold purpose; first, to be used to generate narrowband µJ-level 2.05 µm pulses for seeding a Ho:YLF amplifier through dual-chirp OPA [17]; second, to be used to generate a CEP stable seed for the broadband LWIR OPCPA through hollow-core fiber broadening and intrapulse difference frequency generation (IDFG) [18]; and third, to be added to the LWIR pulse for two-color HHG thereby enhancing the HHG yield. Since the entire system is driven by a single Ti:Sapphire laser system as a frontend, the final output pulse can be CEP stabilized. Figure 1 shows the schematic of our proposed two-color HHG scheme. Furthermore, it is advantageous to seed the Ho:YLF pump laser with µJ-level pulses as the overall gain requirement reduces by 3 - 4 orders of magnitude as compared to that when seeded by nJ-level pulse, thereby reducing the gain narrowing of the amplifier. Figure 2 shows the schematic of the 2-µm pulse generation with the Ti:Sapphire laser as frontend.

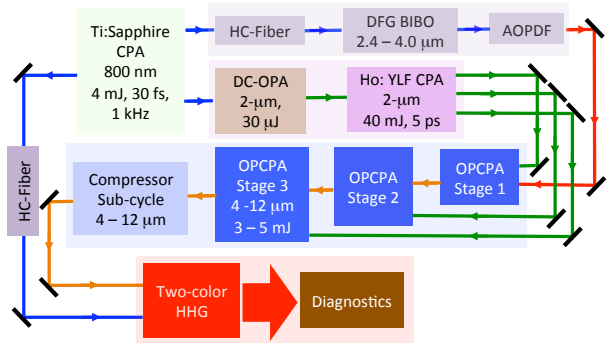

**Fig.1.** Schematic of the two-color HHG driven by LWIR with Ti:Sapphire laser as frontend.

Here we report on a simple, efficient and robust Ho:YLF based multi-stage amplifier at a 1 kHz repetition rate and picosecond pulse duration. We demonstrate amplification of µJ-level pulses to 36 mJ in a four stage amplifier with CW pumping at an extraction efficiency of ~14%. The Ho:YLF crystals are end-pumped using three commercial CW Tm-fiber lasers from IPG Photonics Inc. The amplifier employs a conventional chirped pulse amplification scheme. A single reflection-based Chirp Volume Bragg Grating (CVBG) is used for both stretching and compression of the pulses, which significantly reduces the overall system footprint as compared with a system that utilizes diffraction gratings. Typically such grating (600 l/mm) compressor has a footprint of around 180 cm x 70 cm [14] while the footprint of CVBG is few centimeters. Multiple amplification stages are deployed in our system, which allows us to easily optimize the spot size of the seed beam independently in the different gain crystals. The laser system is maintained at room temperature while the crystals are water-cooled to 18$^o$C and is purged with dry nitrogen to a humidity level of < 1%.

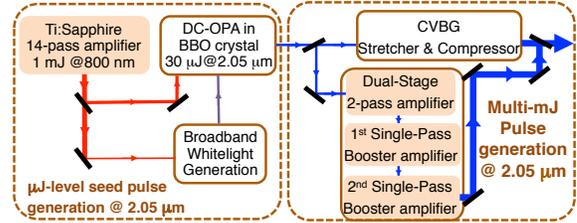

**Fig.2.** Schematic of the 2-µm pulse generation.

The seed source consists of NIR DC-OPA which is pumped by a home-built 14-pass Ti:Sapphire amplifier that emits pulses at 30 fs and 4 mJ at a 1 kHz repetition rate. A 3-mJ pulse centered at 785 nm is compressed to single-cycle by a gas filled hollow-core fiber. Only 1 mJ pulse is used to pump the DC-OPA, which is divided into two beams using a 1% reflection and 99% transmission beam splitter. The weaker of the two beams is focused in a 3 mm YAG window to generate broadband white light. The white light is then positively chirped before being used to seed a single-stage BBO based collinear OPA, which is pumped by the remaining of 0.99 mJ. The OPA generates 36 µJ idler pulses at 2051 nm. The details of the OPA can be found here [17]. The idler pulse at 2051 nm is chosen to match the gain bandwidth of Ho:YLF crystal and has a FWHM bandwidth of 20 nm. However, due to the limited reflection bandwidth of the CVBG in the stretcher, the idler spectrum bandwidth is further reduced to ~4.5 nm leading to a pulse energy of 2 µJ. It must be noted here that it is not advantageous to seed the amplifier with a broader spectrum as the limited gain bandwidth of the Ho:YLF crystal that will lead to a narrowed spectrum after amplification. Figure 3 shows the schematic design of the multi-stage amplifier. The seed pulses are incident normally on the CVBG from the red-to-blue side in the forward direction through the polarization-rotation scheme that utilizes a thin-film polarizer (TFP) and a quarter waveplate (QWP). The seed beam is stretched to pulse duration of 490 ps and is reflected backwards. The stretched beam is reflected off the same TFP due to the rotation from *p*- to *s*-polarization, which is then injected into the first dual-stage 2-pass amplifier. Before injection, the beam is passed through a Faraday Rotator (FR). The FR rotates the incoming *s*-polarized seed beam by 45$^0$ in the forward direction and the half-wave plate (HWP) is placed such that it further rotates the polarization by 45$^0$ thereby converting the seed beam from *s*- to *p*-polarization and is seeded in the first and second stage amplifier.

Two identical gain crystals are used in the first and second stages. The crystals are 50 mm long, a-cut, 0.75% doped Ho:YLF cylindrical rods with a clear apertures of 5 mm diameter. The rods are mounted on indium-contacted copper holders that are water-cooled to 18$^o$C. The crystals are pumped using a single Tm-fiber laser that emits unpolarized continuous wave (CW) 1940 nm light with a power of 120 W. The absorption cross-sections of the pump for polarizations perpendicular and parallel to the c-axis are 0.58 x 10$^{-20}$ cm$^2$ and 1 x 10$^{-20}$ cm$^2$ respectively, while the emission cross-sections at 2051 nm are 0.78 x 10$^{-20}$ cm$^2$ and 1.5 x 10$^{-20}$ cm$^2$ respectively. The polarizations of both pump and seed beams are parallel to the crystal c-axis. The optical axis of the crystal is aligned by rotating the crystal with the incident pump beam and optimized for maximum absorption before it is mounted in the copper holder. The non-saturated pump absorption in the crystal is 99%.

The CW output of the Tm-fiber laser is first split into two equal *s*- and *p*-polarized beam using an optically contacted polarizing beam splitter (PBS) cube leading to 60 W of maximum pump power in each arm of the PBS. In the co-linear geometry, the *p*-polarized transmitted beam through the PBS is sent to the second stage amplifier while the *s*-polarized reflected beam from the PBS is then converted to *p*-polarization by passing it through a half-wave plate (HWP) and then sent into the first stage amplifier. The pump is tightly focused to a beam diameter ($1/e^2$) of 0.7 mm and mode matched in the crystal with the seed using a telescope with a convex and a concave lens combination. The stretched seed pulses are focused into the first stage amplifier crystal using a lens of focal length 400 mm to a beam diameter of 0.7 mm. After amplification of the seed pulse in the first stage, the unabsorbed pump is separated using a dichroic mirror that is highly reflective at 2051 nm and transmits at 1940 nm. The dichroic mirrors are custom dielectrics coated to reflect >99% at 2050 nm for the seed beam while transmitting >98% at 1940 nm of the pump beam. The unabsorbed pump power is then dumped out of the system and the amplified pulse is sent to the second stage amplifier with the same parameters and design geometry as that of the first stage. The remaining pump can potentially be reused in the double-sided pump geometry to improve the amplifier optical efficiency of the system.

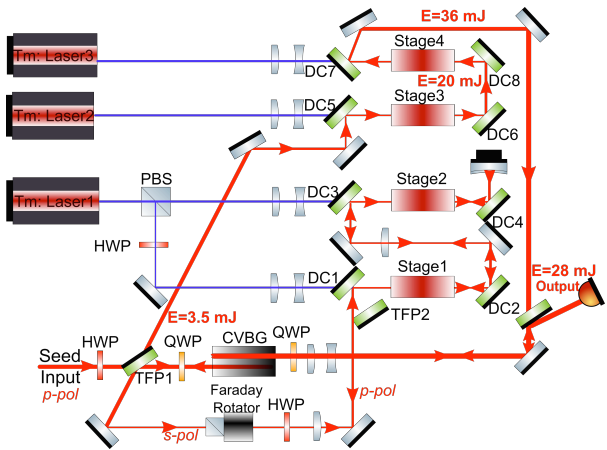

**Fig.3.** Experimental layout of the amplifier system

After amplification of the seed beam in the first stage and second stage amplifier, the seed is then refocused back into the two-stage amplifier using a concave mirror. Since the back reflected beam passes again through the first stage, the radius of curvature of the concave mirror is precisely chosen to keep the seed spot size in the first stage (fourth pass) below the damage threshold of the crystal. After the amplification in the first and second stage, the amplified pulse re-encounters the HWP and FR. However, in the reverse direction due to the directional property of the FR, the effect of HWP is reversed and the exiting pulse remains *p*-polarized. This the *p*-polarized amplified beam is transmitted through the TFP thereby exiting the dual-stage amplifier. Thus the amplified pulse from first and second stage is ejected out of the amplifier using the four-way port of the TFP and the pulse is amplified to 3.5 mJ energy with a total pump power of 28 W. The pulse is subsequently amplified to two single-pass booster amplifiers. The seed fluence in each stage is precisely chosen such that there is a tradeoff between the extraction efficiency and the damage threshold of the AR coating of the crystal.

The third stage amplifier is aligned similarly to the first and second stage. However in this stage the seed pulses passes through the gain medium just once. The crystal length is 50 mm. The pump and seed beam diameters are maintained at 1.5 mm. The pump power is derived from a second Tm-fiber laser with a total unpolarized pump power of 120 W. Both the pump and seed beams are aligned collinearly into the gain medium by combining through the dichroic mirror DC5. However, in order to operate the dichroic mirror DC5 for both an unpolarized pump beam at 1940 nm with transmission >98% and the *p*-polarized seed beam at 2051 nm with reflection >99%, the dielectric coating on dichroic mirror DC5 is redesigned for an incident angle (AOI) <$20^0$ contrary to that used in the previous stages with AOI at $45^0$. After amplification the amplified and the unabsorbed pump beams are again separated using a second dichroic mirror DC6. The unabsorbed pump beam in the Ho:YLF crystal is transmitted out to a beam dump while the amplified beam is sent into the fourth stage. The 3.5-mJ pulses from the second stage are then amplified to 20 mJ in the third stage by an unpolarized pump power of 120 W. Figure 4 shows the output energy from the first three-stage amplifier plotted against the pump power.

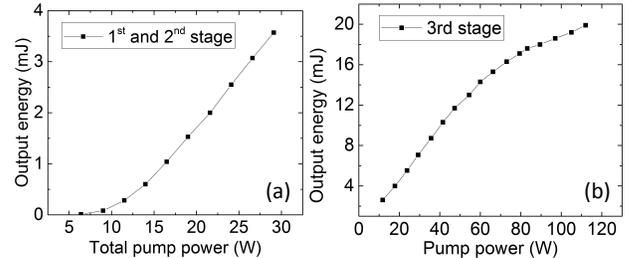

**Fig.4.** Output energy vs. pump power (a) First and second stage output energy (b) Third stage output energy.

Finally, the output of the third stage seeds the fourth stage for further amplification. The fourth stage amplifier is pumped by unpolarized beam of 130 W. The dichroic mirror DC7 combines the pump and seed beam with AOI of <$20^0$ similar to DC5. The pump and seed beam diameters in the fourth stage are chosen to be 2.2 mm to avoid laser damage. After amplification, the unabsorbed pump and the amplified seed beam are then separated using another dichroic mirror DC8. Due to the lower pump absorption for the polarization perpendicular to the c-axis of the crystal, the transmitted power is high. Therefore, a water-cooled beam dump is used to reduce the heating effect on the dump. In the fourth stage the 20-mJ seed pulse is further amplified to 36 mJ. Although reducing the seed beam diameter permits us to increase the amplification beyond 40 mJ but the pulse energy was restricted to operate in the safe regime away of damage threshold of the AR coating to avoid damage. Due to the higher dn/dt along c-axis of the crystal, the thermal lensing effect stretches the beam profile in one direction. Therefore the 4$^{th}$ stage crystal is rotated by 90-degree respect to the 3$^{rd}$ stage in order to balance the astigmatism. The polarization of the seed is rotated accordingly to ensure the optimum amplification. However, the output beam profile still shows some degree of ellipticity at the far field.

Finally, the amplified pulse is compressed using the same CVBG that was used initially for stretching. The front side of the CVBG was used to stretch the pulse from the red-to-blue side while the backside is used to compress the pulse from the blue-to-red side. This configuration avoids any chirp mismatch arising due to possible misalignment where separate gratings are used for stretching and compression and is economical avoiding the cost of an extra pair of gratings. Before sending the amplified pulses to the CVBG for compression, the pulses are sent through the similar configuration of TFP and QWP as used before in the stretcher. The amplified pulses are sent into the CVBG at normal incidence. Figure 5 shows the results obtained after amplification from the fourth stage. Figure 5(a) shows the output energy plot after the fourth stage before (black) and after compression (red) vs total pump power used in all the stages. The transverse dimension of the CVBG is 27 x 20 mm$^2$ clear aperture and length of 49 mm with a spectral bandwidth of 4.5 nm. The CVBG are specially designed to operate at high energy by Optigrate Inc. and can operate above a peak power of 3 GW/cm$^2$ and pulse energy above 50 mJ. With the 36-mJ input on to the CVBG, the output energy is 28 mJ yielding the compression efficiency of 78%. This is the highest pulse energy ever achieved in a kHz Ho:YLF laser system utilizing CVBG for compressor. The previous compression reported in CVBG was operated below 25 mJ to avoid the damage of the CVBG [15]. In addition our autocorrelation results show reduced energy content in the satellites. Figure 5 (b) shows the output spectrum measured using a high resolution of (0.85 nm) NIRQuest spectrometer from Ocean Optics. $\lambda_{FWHM}$ of the spectrum of 2.2 nm is achieved. Figure 5(c) shows the beam profile after the fourth stage before compression measured using Spiricon camera. To characterize the temporal profile of the final amplified and compressed pulse, a commercial autocorrelator for 2 μm is used. The autocorrelation trace is shown in the Figure 5(d) with a sech$^2$-fiting showing 10 ps autocorrelation full-width at half maxima (FWHM) corresponding to a FWHM pulse duration of 6.8 ps. Compared to the results reported in Ref. 15 and Ref. 16, our pulse does not have any satellite pulse. We believe this might be due to the absence of the Pockels Cell in our design contrary to that used in their RA.

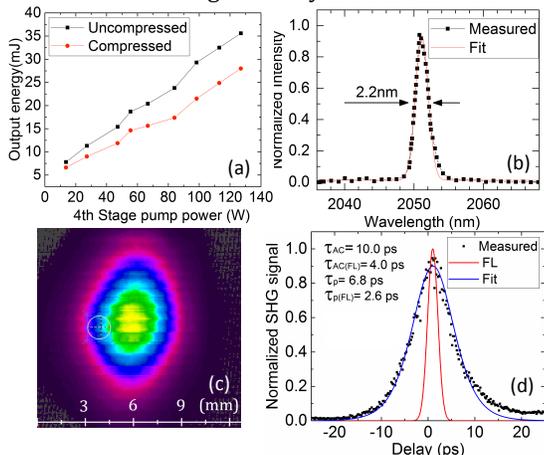

**Fig.5.** Experimental results after fourth stage (a) Output energy before (black) and after compression (red) vs. total pump power used in all the stages, (b) Output spectrum $\lambda_{FWHM}$ = 2.2 nm (c) Beam profile before compression, (d) The autocorrelation trace with sech$^2$-fitting; $\tau_{ac}$ = 10 ps is the autocorrelation duration, $\tau_p$ = 6.8 ps is the fitted pulse duration.

In conclusion we have developed a compact high power Ho:YLF amplifier at 1 kHz with a pulse energy > 28 mJ. The pulses are stretched and compressed using a single CVBG, with the output compressed to $\tau_p$ = 6.8 ps. The final output energy can be further increased by utilizing the full pump power in the first two stages and will be used to pump sub-cycle LWIR OPCPAs [19]. The amplifier is seeded by a Ti:Sapphire frontend that paved the way for two-color driven HHG to generate high flux keV attosecond X-rays.

**Funding.** U.S Air force Office of Scientific Research (AFSOR) (FA9550-17-1-0499, FA9550-16-1-0013, and FA9550-15-1-0037), Army Research Office (ARO) (W911NF-19-1-0224 and W911NF-14-1-0383), Defense Advanced Research Project Agency (DARPA) (D18AC00011), Defense Threat Reduction Agency (HDTRA11910026), National Science Foundation (NSF) (1806575), DSTL/EPSRC (MURI EP/N018680/1).

**Disclosures.** The authors declare no conflicts of interest.

**References**

1. X. Ren, J. Li, Y. Yin, K. Zhao, A. Chew, Y. Wang, S. Hu, Y. Cheng, E. Cunningham, Y. Wu, M. Chini, and Z. Chang, J. Opt. **20**, 023001 (2018).
2. G. J. Stein, P. D. Keathley, P. Krogen, H. Liang, J. P. Siqueira, C.-L. Chang, C.-J. Lai, K.-H. Hong, G. M. Laurent, and F. X. Kärtner, J. Phys. B At. Mol. Opt. Phys. **49**, 155601 (2016).
3. S. Ghimire, A. D. DiChiara, E. Sistrunk, P. Agostini, L. F. DiMauro, and D. A. Reis, Nat. Phys. **7**, 138–141 (2011).
4. I. Jovanovic, G. Xu, and S. Wandel, Phys. Procedia **52**, 68–74 (2014).
5. H. R. Reiss, Phys. Rev. Lett. **101**, 043002 (2008).
6. S. Woutersen, Science **278**, 658–660 (1997).
7. T. Popmintchev, M.-C. Chen, D. Popmintchev, P. Arpin, S. Brown, S. Alisauskas, G. Andriukaitis, T. Balciunas, O. D. Mucke, A. Pugzlys, A. Baltuska, B. Shim, S. E. Schrauth, A. Gaeta, C. Hernandez-Garcia, L. Plaja, A. Becker, A. Jaron-Becker, M. M. Murnane, and H. C. Kapteyn, Science **336**, 1287–1291 (2012).
8. B. Shan and Z. Chang, Phys. Rev. A **65**, 011804 (2001).
9. G. Orlando, P. P. Corso, E. Fiordilino, and F. Persico, J. Phys. B At. Mol. Opt. Phys. **43**, 025602 (2010).
10. M. B. Gaarde, K. J. Schafer, A. Heinrich, J. Biegert, and U. Keller, Phys. Rev. A - At. Mol. Opt. Phys. **72**, (2005).
11. Z. Chang, OSA Contin. **2**, 2131 (2019).
12. K. Murari, H. Cankaya, P. Kroetz, G. Cirmi, P. Li, A. Ruehl, I. Hartl, and F. X. Kärtner, Opt. Lett. **41**, 1114 (2016).
13. P. Kroetz, A. Ruehl, G. Chatterjee, A.-L. Calendron, K. Murari, H. Cankaya, P. Li, F. X. Kärtner, I. Hartl, and R. J. Dwayne Miller, Opt. Lett. **40**, 5427 (2015).
14. M. Hemmer, D. Sánchez, M. Jelínek, V. Smirnov, H. Jelinkova, V. Kubeček, and J. Biegert, Opt. Lett. **40**, 451–454 (2015).
15. L. von Grafenstein, M. Bock, D. Ueberschaer, U. Griebner, and T. Elsaesser, Opt Lett **41**, 4668–4671 (2016).
16. M. Bock, L. von Grafenstein, U. Griebner, and T. Elsaesser, J. Opt. Soc. Am. B **35**, C18 (2018).
17. Y. Yin, X. Ren, Y. Wang, F. Zhuang, J. Li, and Z. Chang, Photonics Res. **6**, 1 (2018).
18. Y. Yin, X. Ren, A. Chew, J. Li, Y. Wang, F. Zhuang, Y. Wu, and Z. Chang, Sci. Rep. **7**, 11097 (2017).
19. L. von Grafenstein, M. Bock, D. Ueberschaer, U. Griebner, and T. Elsaesser, Opt. Lett. **41**, 4668 (2016).
20. Y. Yin, A. Chew, X. Ren, J. Li, Y. Wang, Y. Wu, and Z. Chang, Sci. Rep. **8**, 45794 (2017).

**References**

1. X. Ren, J. Li, Y. Yin, K. Zhao, A. Chew, Y. Wang, S. Hu, Y. Cheng, E. Cunningham, Y. Wu, M. Chini, and Z. Chang, "Attosecond light sources in the water window," J. Opt. **20**, 023001 (2018).
2. G. J. Stein, P. D. Keathley, P. Krogen, H. Liang, J. P. Siqueira, C.-L. Chang, C.-J. Lai, K.-H. Hong, G. M. Laurent, and F. X. Kärtner, "Water-window soft x-ray high-harmonic generation up to the nitrogen K-edge driven by a kHz, 2.1 μm OPCPA source," J. Phys. B at. Mol. Opt. Phys. **49**, 155601 (2016).
3. S. Ghimire, A. D. DiChiara, E. Sistrunk, P. Agostini, L. F. DiMauro, and D. A. Reis, "Observation of high-order harmonic generation in a bulk crystal," Nat. Phys. **7**, 138–141 (2011).
4. I. Jovanovic, G. Xu, and S. Wandel, "Mid-infrared laser system development for dielectric laser accelerators," Phys. Procedia **52**, 68–74 (2014).
5. H. R. Reiss, "Limits on Tunneling Theories of Strong-Field Ionization," Phys. Rev. Lett. **101**, 043002 (2008).
6. S. Woutersen, "Femtosecond Mid-IR Pump-Probe Spectroscopy of Liquid Water: Evidence for a Two-Component Structure," Science **278**, 658–660 (1997).
7. T. Popmintchev, M.-C. Chen, D. Popmintchev, P. Arpin, S. Brown, S. Alisauskas, G. Andriukaitis, T. Balciunas, O. D. Mucke, A. Pugzlys, A. Baltuska, B. Shim, S. E. Schrauth, A. Gaeta, C. Hernandez-Garcia, L. Plaja, A. Becker, A. Jaron-Becker, M. M. Murnane, and H. C. Kapteyn, "Bright Coherent Ultrahigh Harmonics in the keV X-ray Regime from Mid-Infrared Femtosecond Lasers," Science **336**, 1287–1291 (2012).
8. B. Shan and Z. Chang, "Dramatic extension of the high-order harmonic cutoff by using a long-wavelength driving field," Phys. Rev. A **65**, 011804 (2001).
9. G. Orlando, P. P. Corso, E. Fiordilino, and F. Persico, "A three-colour scheme to generate isolated attosecond pulses," J. Phys. B At. Mol. Opt. Phys. **43**, 025602 (2010).
10. M. B. Gaarde, K. J. Schafer, A. Heinrich, J. Biegert, and U. Keller, "Large enhancement of macroscopic yield in attosecond pulse train assisted harmonic generation," Phys. Rev. A - At. Mol. Opt. Phys. **72**, (2005).
11. Z. Chang, "Enhancing keV high harmonic signals generated by long-wave infrared lasers," OSA Contin. **2**, 2131 (2019).
12. K. Murari, H. Cankaya, P. Kroetz, G. Cirmi, P. Li, A. Ruehl, I. Hartl, and F. X. Kärtner, "Intracavity gain shaping in millijoule-level, high gain Ho:YLF regenerative amplifiers," Opt. Lett. **41**, 1114 (2016).
13. P. Kroetz, A. Ruehl, G. Chatterjee, A.-L. Calendron, K. Murari, H. Cankaya, P. Li, F. X. Kärtner, I. Hartl, and R. J. Dwayne Miller, "Overcoming bifurcation instability in high-repetition-rate Ho:YLF regenerative amplifiers," Opt. Lett. **40**, 5427 (2015).
14. M. Hemmer, D. Sánchez, M. Jelínek, V. Smirnov, H. Jelinkova, V. Kubeček, and J. Biegert, "2-μm wavelength, high-energy Ho:YLF chirped-pulse amplifier for mid-infrared OPCPA," Opt. Lett. **40**, 451–454 (2015).
15. L. von Grafenstein, M. Bock, D. Ueberschaer, U. Griebner, and T. Elsaesser, "Ho:YLF chirped pulse amplification at kilohertz repetition rates – 4.3 ps pulses at 2 μm with GW peak power," Opt. Lett. **41**, 4668 (2016).
16. M. Bock, L. von Grafenstein, U. Griebner, and T. Elsaesser, "Generation of millijoule few-cycle pulses at 5 μm by indirect spectral shaping of the idler in an optical parametric chirped pulse amplifier," J. Opt. Soc. Am. B **35**, C18 (2018).
17. Y. Yin, X. Ren, Y. Wang, F. Zhuang, J. Li, and Z. Chang, "Generation of high-energy narrowband 2.05 μm pulses for seeding a Ho:YLF laser," Photonics Res. **6**, 1 (2018).
18. Y. Yin, X. Ren, A. Chew, J. Li, Y. Wang, F. Zhuang, Y. Wu, and Z. Chang, "Generation of octave-spanning mid-infrared pulses from cascaded second-order nonlinear processes in a single crystal," Sci. Rep. **7**, 11097 (2017).
19. Y. Yin, A. Chew, X. Ren, J. Li, Y. Wang, Y. Wu, and Z. Chang, "Towards Terawatt Sub-Cycle Long-Wave Infrared Pulses via Chirped Optical Parametric Amplification and Indirect Pulse Shaping," Sci. Rep. **8**, 45794 (2017).